\documentclass[english,aps, superscriptaddress, floatfix, 12pt]{revtex4}
\usepackage[T1]{fontenc}
\usepackage[latin1]{inputenc}
\usepackage{graphicx}
\usepackage{amssymb}
\usepackage{epsfig}
\usepackage{amsmath}
\usepackage{psfrag}
\usepackage[active]{srcltx}
\usepackage{babel}
\def\be{\begin{equation}}
\def\ee{\end{equation}}
\def\bea{\begin{eqnarray}}
\def\eea{\end{eqnarray}}
\makeatother   
\begin{document}
\title{Triviality of the Aharonov--Bohm interaction in a spatially confining vacuum}
\author{Dmitri Antonov\\
{\it Departamento de F\'isica and Centro de F\'isica das Interac\c{c}\~oes Fundamentais,}\\ 
{\it Instituto Superior T\'ecnico, UT Lisboa,
Av. Rovisco Pais, 1049-001 Lisboa, Portugal}}

\noaffiliation

\begin{abstract}
This paper explores long-range interactions between magnetically-charged excitations of the vacuum of the dual
Landau--Ginzburg theory (DLGT) and the dual Abrikosov vortices present in the same vacuum. 
We show that, in the London limit of DLGT,
the corresponding Aharonov--Bohm-type interactions possess such a coupling that the interactions reduce to a trivial
factor of ${\rm e}^{2\pi i\times({\rm integer})}$. The same analysis is done in the SU($N_c$)-inspired 
[U(1)]$^{N_c-1}$-invariant DLGT, as well as in DLGT extended by a Chern--Simons term. It is furthermore 
explicitly shown 
that the Chern--Simons term leads to the appearance of knotted dual Abrikosov vortices.
\end{abstract}

\maketitle

\section{Introduction}

It has long been known that quark confinement 
in QCD can be modeled by means of a dual-superconductor scenario~\cite{1, 15}.
This scenario suggests that the Yang--Mills vacuum can resemble that of a dual superconductor, which consists of the condensate of a magnetically charged Higgs field. The resulting dual Abelian Higgs model is a four-dimensional relativistic generalization 
of the Landau--Ginzburg theory of dual superconductivity. Dedicated lattice simulations support 
this scenario of confinement with a very high accuracy~\cite{2}.

It turns out that not only the dual Abelian Higgs model but also 
the dual Landau--Ginzburg theory (DLGT) can be relevant to the description of the Yang--Mills vacuum.
The reason is that, upon the deconfinement phase transition, large spatially-oriented Wilson loops 
still exhibit an area-law behavior (see Ref.~\cite{f1} for the lattice 
results on the corresponding spatial string tension $\sigma_s$). Analytically, 
spatial confinement can with a good accuracy be  
described in terms of soft stochastic chromo-magnetic Yang--Mills fields~\cite{ag}, which (unlike soft
chromo-electric fields) survive the deconfinement phase transition~\cite{de}.
Moreover, for every temperature-dependent quantity, there exists the so-called temperature of dimensional reduction
such that, above that temperature, the contribution to the quantity at issue produced by all Matsubara frequencies $\omega_k=2\pi Tk$ with $k\ne 0$
is negligible compared to the contribution of $\omega_0$. It should be, of course, borne in mind that, although 
the contributions of nonzero modes amount to at most few per cent of the static-mode 
contribution, these contributions are always present. For this reason, the dimensional reduction is not a phase transition with a definite 
critical temperature that can be determined from the thermodynamic equations. At the formal level, one can only say that
the dimensional reduction of the Euclidean Yang--Mills action corresponds to the substitution
\be
\label{dr}
S_{\rm YM}=\frac{1}{4g_{\rm YM}^2}\int d^3x\int_{0}^{1/T} dx_4{\,}(F_{\mu\nu}^a)^2 \rightarrow
\frac{1}{4g_{\rm YM}^2T}\int d^3x{\,}(F_{\mu\nu}^a)^2,
\ee
where $F_{\mu\nu}^a=\partial_\mu A_\nu^a-\partial_\nu A_\mu^a-f^{abc}A_\mu^b A_\nu^c$ is the Yang--Mills 
field-strength tensor. Thus, the zero-temperature Yang--Mills coupling $g_{\rm YM}$ goes over to the 
temperature-dependent dimensionful coupling $g_T=g_{\rm YM}\sqrt{T}$. The latter defines the parametric
temperature dependence of all the dimensionful nonperturbative quantities upon their dimensional reduction.
In particular, the spatial string tension scales with temperature as~\cite{ag}
$\sigma_s\propto g_T^4$, ensuring spatial confinement in the 
dimensionally-reduced Yang--Mills theory. As such, this theory 
can be modeled by means of DLGT.

The aim of the present paper is to address topological effects that might occur in DLGT.
These effects are related to the long-range interactions between the excitations of the dual-Higgs vacuum, which are described by Wilson loops, 
and the dual (i.e. carrying electric fluxes) Abrikosov vortices~\cite{az}. The latter 
are present in the vacuum as the topologically stable solutions 
to the classical equations of motion~\cite{15}.
{\it A~priori} one can expect the Wilson loops and Abrikosov vortices to interact only by means of massive dual vector bosons. We show that, in addition, a long-range Aharonov--Bohm-type interaction is present, which appears in the 
form of a Gauss' linking number between the contour of a Wilson loop and an Abrikosov vortex.
However, in the so-called London limit, which corresponds to an extreme type-II dual superconductor, 
the coupling of the Aharonov--Bohm-type interaction is shown to be such that the interaction trivializes, 
producing only an inessential factor of ${\rm e}^{2\pi i\times({\rm integer})}$.

The paper is organized as follows. In the next Section, we perform a path-integral 
duality transformation of the Wilson loop,
and explicitly find the said Aharonov--Bohm-type interaction. In Section~III, we generalize these results to the 
case of an SU($N_c$)-inspired [U(1)]$^{N_c-1}$-invariant DLGT.
In Section~IV, we additionally consider 
the effects produced in DLGT by the Chern--Simons (CS) term. First, we briefly show that, in the absence of the dual
Higgs field, the CS term leads to a self-linkage of the contour of the Wilson loop. Then we 
perform the duality transformation of the Wilson loop in the full 
theory, which includes the dual Higgs field. In particular, at sufficiently large values of the $\Theta$-parameter
entering the CS term, we obtain an analytic expression for the Wilson loop. 
Furthermore, in the same large-$\Theta$
limit, we explicitly find knotted dual Abrikosov vortices, whose self-linkage is provided by 
the CS term. In Section~IV, the summary of the results obtained is presented.
In Appendices A and B, we provide some 
technical details of the calculations performed.

\section{Wilson loop in the dual Landau--Ginzburg theory}

Dual Abelian Higgs model is described by the following Euclidean action:
$$S_{\rm DAHM}=\int d^4x\left\{\frac14F_{\mu\nu}^2[B]+|D_\mu \varphi|^2+\lambda(|\varphi|^2-\eta_{\rm 4d}^2)^2\right\}.$$
Here $F_{\mu\nu}[B]=\partial_\mu B_\nu-\partial_\nu B_\mu$ is the strength tensor of the dual gauge field $B_\mu$, and 
$D_\mu=\partial_\mu+ig_mB_\mu$ is the covariant derivative, with $g_m$ being the dimensionless magnetic coupling 
related to the electric coupling $e$ via the Dirac quantization condition $g_m e=2\pi\times{\,}({\rm integer})$. We consider this model in the so-called London limit of $\sqrt{\lambda}\gg g_m$, that is, the extreme 
type-II dual superconductor. Due to the factor ${\rm e}^{-\lambda\int d^4x(|\varphi|^2-\eta_{\rm 4d}^2)^2}$
in the partition function, the dominant contribution to the functional integral is produced by configurations
of the dual Higgs field with $|\varphi|=\eta_{\rm 4d}$. That is, variations of the radial part of the 
dual-Higgs field do not matter in the London limit, which is equivalent to the fact that the condensate of this field is fully developed everywhere except of infinitely thin cores of the dual strings. Rather, it is the phase of the 
dual Higgs field which matters, so that $\varphi(x)=\eta_{\rm 4d}{\,}{\rm e}^{i\theta(x)}$, and the 
kinetic term of the dual Higgs field takes the form $|D_\mu \varphi|^2=\eta_{\rm 4d}^2\cdot(\partial_\mu\theta+
g_mB_\mu)^2$. Accordingly, 
in the London limit of interest, 
the action of the dual Abelian Higgs model reads
\begin{equation}
\label{s4d}
S_{4{\rm d}}=\int d^4x\left\{\frac14F_{\mu\nu}^2[B]+\eta_{4{\rm d}}^2(\partial_\mu\theta+g_m B_\mu)^2\right\}.
\end{equation}
Notice that this action can be used to calculate the tension of a Nambu--Goto string interconnecting two static 
electric charges, as well as the correlation length of the two-point function of $F_{\mu\nu}$'s (cf. Ref.~\cite{ae1}).
Matching these two quantities with their phenomenological QCD-counterparts, one can readily find
$\eta_{4{\rm d}}\sim\sqrt{\sigma}$ and $g_m\sim\frac{1}{a\sqrt{\sigma}}$,
where $\sigma$ is the string tension entering the static quark-antiquark potential, and $a$ is the 
correlation length of the two-point correlation function of gluonic field strengths.

As was mentioned in Introduction, upon the deconfinement phase transition in QCD, the chromo-electric part of the 
gluon condensate vanishes (in accordance with deconfinement), while the chromo-magnetic part survives, providing 
an area law for large spatial Wilson loops (cf. Refs.~\cite{f1,ag,de}). The corresponding spatially confining 
vacuum can be modelled by means of the dual Landau--Ginzburg theory.
The action of this theory,
\begin{equation}
\label{s3dd}
S_{3{\rm d}}=\int d^3x\left\{\frac{1}{4}F_{\mu\nu}^2[b]+
\eta_{3{\rm d}}^2(\partial_\mu\theta+\kappa b_\mu)^2\right\},
\end{equation}
follows from the action~(\ref{s4d}) upon the substitution $\int d^4x\to\beta\int d^3x$, where $\beta\equiv1/T$ [cf.
the same substitution in the Yang--Mills action~(\ref{dr})].
Matching the fields and parameters of the action $S_{3{\rm d}}$ with those of the action $S_{4{\rm d}}$, we obtain the following relations:
\be
\label{rel}
b_\mu=\sqrt{\beta}B_\mu,~~ \eta_{3{\rm d}}=\sqrt{\beta}\eta_{4{\rm d}},~~ \kappa=g_m\sqrt{T}.
\ee
Notice that, in terms of the phenomenological QCD parameters $\sigma$ and $a$ (cf. the previous paragraph), one gets the estimates $\eta_{3{\rm d}}\sim\sqrt{\sigma\beta}$, $\kappa\sim\frac{\sqrt{T/\sigma}}{a}$.

We consider now the central object of our study, that is, the Wilson loop associated with an excitation of 
the dual-Higgs vacuum. In the initial dual Abelian Higgs model, it has the form 
$\left<W(C)\right>_{\rm DAHM}=\left<\exp\left(ig_mN\oint_C dx_\mu B_\mu\right)\right>$, where the integer $N$ 
characterizes the magnetic charge $g_mN$ of an excitation that propagates along the contour $C$.
The counterpart of this expression in the dual Landau--Ginzburg theory reads
\be
\label{we}
\left<W(C)\right>=\left<\exp\left(i\kappa N\oint_C dx_\mu b_\mu\right)\right>,
\ee 
where we have used the 
above relations~(\ref{rel}). We notice that, in the purely Maxwell theory
corresponding to $\eta_{3{\rm d}}=0$ in Eq.~(\ref{s3dd}), the Wilson loop has the form 
\be
\label{wt}
\left<W(C)\right>=\exp\left(-\frac{(\kappa N)^2}{2}
\oint_C dx_\mu \oint_C dy_\mu D_0({\bf x}-{\bf y})\right),
\ee 
where $D_0({\bf x})=1/(4\pi|{\bf x}|)$ is the Coulomb propagator.

We calculate now the Wilson loop $\left<W(C)\right>$ with the average $\left<\cdots\right>$ corresponding to the full action~(\ref{s3dd}), where $\eta_{3{\rm d}}\ne 0$. To this end, we find it convenient to introduce, instead of the 
field $b_\mu$, a rescaled field $v_\mu=b_\mu/(\kappa N)$, and denote
\be
\label{numu}
\nu=1/(\kappa N)^2,~~~~~ \mu=\kappa^2N.
\ee 
In terms of these notations, the Wilson loop~(\ref{we})
can be written as
\begin{equation}
\label{1}
\left<W(C)\right>=\int {\cal D}v_\mu{\,}{\cal D}\tilde\theta{\,}{\cal D}\bar\theta{\,}
{\rm e}^{-\int_x\left[\frac{1}{4\nu}F_{\mu\nu}^2[v]+\eta^2(\partial_\mu\theta+\mu v_\mu)^2-
\frac{i}{\nu}v_\mu j_\mu\right]},
\end{equation}
where $j_\mu({\bf x};C)=\oint_C dx_\mu(\tau)\delta({\bf x}-
{\bf x}(\tau))$ is a conserved current, $\eta\equiv\eta_{3{\rm d}}$, and 
from now on we use the short-hand notations $\int_x\equiv\int d^3x$ and $\int_p\equiv \int \frac{d^3p}{(2\pi)^3}$.
The full phase $\theta$ of the dual Higgs field can be represented as a sum $\theta=\tilde\theta+\bar\theta$,
with $\tilde\theta$ experiencing jumps by $2\pi$ when going around dual Abrikosov vortices,
while $\bar\theta$ being a Gaussian fluctuation around $\tilde\theta$. 
The said jumps of $\tilde\theta$
lead to the noncommutativity of two derivatives acting on this field (cf. Ref.~\cite{15}): 
\begin{equation}
\label{2}
(\partial_\mu\partial_\nu-\partial_\nu\partial_\mu)\tilde\theta=2\pi\varepsilon_{\mu\nu\lambda}J_\lambda,
\end{equation}
where $J_\lambda$ is a current of the dual Abrikosov vortex.

To calculate the Wilson loop~(\ref{1}), we perform its duality transformation.
To this end, 
it is first convenient to introduce two auxiliary fields as follows:
$${\rm e}^{-\frac{1}{4\nu}\int_x F_{\mu\nu}^2}=
\int {\cal D}G_\mu{\,}{\rm e}^{\int_x\left[-\frac{\nu}{2}G_\mu^2+i\varepsilon_{\mu\nu\lambda}v_\mu\partial_\nu G_\lambda\right]},~
{\rm e}^{-\eta^2\int_x(\partial_\mu\theta+\mu v_\mu)^2}=\int {\cal D}C_\mu{\,} {\rm e}^{\int_x\left[
-\frac{1}{4\eta^2}C_\mu^2+iC_\mu(\partial_\mu\theta+\mu v_\mu)\right]}.$$
The subsequent integration over $\bar\theta$ leads to the constraint $\partial_\mu C_\mu=0$, which can be 
resolved by representing $C_\mu$ as $C_\mu=\varepsilon_{\mu\nu\lambda}\partial_\nu\varphi_\lambda$.
Accordingly,
$C_\mu^2=\frac12\Phi_{\mu\nu}^2$, where $\Phi_{\mu\nu}=\partial_\mu\varphi_\nu-\partial_\nu\varphi_\mu$, and 
$i\int_x C_\mu\partial_\mu\tilde\theta=2\pi i\int_x\varphi_\mu J_\mu$, where at the last step we have used 
Eq.~(\ref{2}). Thus, the Wilson loop~(\ref{1}) takes the form
\begin{equation}
\label{qq}
\left<W(C)\right>=\int {\cal D}J_\mu{\,}{\cal D}\varphi_\mu{\,}{\cal D}G_\mu{\,}
{\cal D}v_\mu{\,}
{\rm e}^{\int_x\left[-\frac{\nu}{2}G_\mu^2-\frac{1}{8\eta^2}\Phi_{\mu\nu}^2+i\varepsilon_{\mu\nu\lambda}
v_\mu\partial_\nu(G_\lambda+\mu\varphi_\lambda)+2\pi i\varphi_\mu J_\mu+\frac{i}{\nu}v_\mu j_\mu\right]}.
\end{equation}
Note that, throughout this paper, we work at the entirely classical level. For this reason, the 
Jacobian corresponding to the change of integration variables $\tilde\theta\to J_\mu$ is 
omitted, and the measure ${\cal D}J_\mu$ in the functional integral has only a statistical (rather than a field-theoretical) meaning of counting vortices in their given configuration.

Next, noticing that the $v_\mu$-field enters 
Eq.~(\ref{qq}) as just a Lagrange multiplier, and integrating over this field, we obtain a functional 
$\delta$-function $\delta\left(\varepsilon_{\mu\nu\lambda}\partial_\nu(G_\lambda+\mu\varphi_\lambda)
+\frac{1}{\nu}j_\mu\right)$. The subsequent $G_\mu$-integration amounts to substituting $G_\mu$, which stems from this 
$\delta$-function, into ${\rm e}^{-\frac{\nu}{2}\int_x G_\mu^2}$. Such a $G_\mu$ reads 
$G_\mu=-\mu\varphi_\mu-\frac{1}{\nu}\varepsilon_{\mu\nu\lambda}\int_y\partial_\nu^xD_0^{xy}j_\lambda^y$,
where we have introduced short-hand notations 
$D_0^{xy}\equiv 1/(4\pi|{\bf x}-{\bf y}|)$, $j_\lambda^y\equiv j_\lambda({\bf y};C)$, and used the 
conservation of $j_\mu$. Accordingly, the Wilson loop takes the form 
\be
\label{w5}
\left<W(C)\right>=\int {\cal D}J_\mu{\,}
{\cal D}\varphi_\mu{\,}{\rm e}^{\int_x\left[-\frac{1}{8\eta^2}\Phi_{\mu\nu}^2+2\pi i\varphi_\mu J_\mu
-\frac{\nu}{2}\left(\mu\varphi_\mu+
\frac{1}{\nu}\varepsilon_{\mu\nu\lambda}\int_y\partial_\nu^xD_0^{xy}j_\lambda^y\right)^2\right]},
\ee
or, equivalently,
$$\left<W(C)\right>={\rm e}^{-\frac{1}{2\nu}\int_{x,y}j_\mu^xj_\mu^yD_0^{xy}}
\int {\cal D}J_\mu{\,}
{\cal D}\varphi_\mu{\,}{\rm e}^{\int_x\bigl(-\frac{1}{8\eta^2}\Phi_{\mu\nu}^2-\frac{\mu^2\nu}{2}
\varphi_\mu^2+i\varphi_\mu K_\mu\bigr)},$$
where 
\begin{equation}
\label{k33}
K_\mu^x\equiv 2\pi J_\mu^x+i\mu\varepsilon_{\mu\nu\lambda}\int_y\partial_\nu^xD_0^{xy}j_\lambda^y.
\end{equation}
To perform the remaining $\varphi_\mu$-integration, we introduce a rescaled field $\chi_\mu\equiv\varphi_\mu/(\eta\sqrt{2})$ and denote 
\begin{equation}
\label{mml}
{\sf m}\equiv\mu\eta\sqrt{2\nu}.
\end{equation}
That yields
$$\int {\cal D}\chi_\mu{\,}{\rm e}^{\int_x\bigl[-\frac14(\partial_\mu\chi_\nu-\partial_\nu\chi_\mu)^2-
\frac{{\sf m}^2}{2}\chi_\mu^2+i\sqrt{2}\eta\chi_\mu K_\mu\bigr]}={\rm e}^{-\eta^2\int_{x,y}K_\mu^xK_\mu^y
D_{\sf m}^{xy}},$$
where $D_{\sf m}^{xy}\equiv{\rm e}^{-{\sf m}|{\bf x}-{\bf y}|}/(4\pi|{\bf x}-{\bf y}|)$ is the Yukawa 
propagator. Thus, the Wilson loop~(\ref{we}) becomes
$$\left<W(C)\right>={\rm e}^{-\frac{1}{2\nu}\int_{x,y}j_\mu^xj_\mu^yD_0^{xy}}
\int {\cal D}J_\mu{\,}{\rm e}^{-\eta^2\int_{x,y}K_\mu^xK_\mu^y
D_{\sf m}^{xy}}.$$
The expression standing in the last exponential 
in this formula can be simplified (see Appendix~A for the details), that yields the following result:
\begin{equation}
\label{a2}
\left<W(C)\right>={\rm e}^{-\frac{1}{2\nu}\int_{x,y}j_\mu^xj_\mu^yD_{\sf m}^{xy}}
\int {\cal D}J_\mu{\,}{\rm e}^{-(2\pi\eta)^2\int_{x,y}J_\mu^xJ_\mu^yD_{\sf m}^{xy}+
\frac{2\pi i}{\mu\nu}\left[\hat L(j,J)-\varepsilon_{\mu\nu\lambda}
\int_{x,y}J_\mu^x j_\nu^y\partial_\lambda^xD_{\sf m}^{xy}\right]},
\end{equation}
where $\hat L(j,J)=\varepsilon_{\mu\nu\lambda}\int_{x,y}J_\mu^xj_\nu^y\partial_\lambda^xD_0^{xy}$ is the Gauss'
linking number of the contour $C$ and a dual Abrikosov vortex.
The exponential ${\rm e}^{\frac{2\pi i}{\mu\nu}\hat L(j,J)}$ in Eq.~(\ref{a2}) formally describes a long-range Aharonov--Bohm-type interaction of the dual-Higgs excitation with the dual Abrikosov vortex.
However, recalling the notations introduced in Eq.~(\ref{numu}), we have $\frac{1}{\mu\nu}=N$.
For this reason, the obtained interaction
turns out to be trivial, i.e. ${\rm e}^{\frac{2\pi i}{\mu\nu}\hat L(j,J)}= 1$.
Thus, we conclude that integer-charged excitations of the dual-Higgs vacuum do not interact with the dual Abrikosov vortices by means of the long-range Aharonov--Bohm-type interaction. Rather, the interaction between the 
excitations of the dual-Higgs vacuum and the dual Abrikosov vortices is provided by the dual vector boson,
through the factor ${\rm e}^{-2\pi iN\varepsilon_{\mu\nu\lambda}
\int_{x,y}J_\mu^x j_\nu^y\partial_\lambda^xD_{\sf m}^{xy}}$.

\section{Generalization to the SU($N_c$)-inspired case}

In this Section, we generalize the result~(\ref{a2}) to the SU($N_c$)-inspired case.
The corresponding theory~\cite{ae,s3} is invariant under the $[U(1)]^{N_c-1}$-group, which is the maximal Abelian 
subgroup of SU($N_c$). 
A counterpart of Eq.~(\ref{1}) in this theory reads 
\be
\label{w22}
\left<W_b(C)\right>=\int
{\cal D}{\bf v}_\mu\left(\prod\limits_a {\cal D}\tilde\theta_a{\,}{\cal D}\bar\theta_a\right)
{\cal D}k{\,}\delta\left(\sum\limits_{a}\tilde\theta_a\right)
{\rm e}^{-\int_x\left[\frac{1}{4\nu}{\bf F}_{\mu\nu}^2+\eta^2\sum\limits_a(\partial_\mu\theta_a+\mu 
{\bf q}_a{\bf v}_\mu)^2-ik\sum\limits_a\bar\theta_a-
\frac{i}{\nu}{\bf v}_\mu {\bf j}_\mu^b\right]}.
\ee
Here ${\bf v}_\mu=(v_\mu^1,\ldots,v_\mu^{N_c-1})$,
the index $a=1,\ldots,\frac{N_c(N_c-1)}{2}$ labels positive roots ${\bf q}_a$'s of the SU($N_c$)-group, 
and the fact that this group is special imposes a constraint 
$\sum\limits_{a}^{}\theta_a=0$ on the phases 
$\theta_a$'s of the dual Higgs fields. Similarly to Eq.~(\ref{2}), we have 
$\theta_a=\tilde\theta_a+\bar\theta_a$, where 
$(\partial_\mu\partial_\nu-\partial_\nu\partial_\mu)\tilde\theta_a=2\pi J_\mu^a$, with $J_\mu^a$ being a
current of the dual Abrikosov vortex of the $a$-th type. The constraint $\sum\limits_{a}\bar\theta_a=0$
is further imposed in Eq.~(\ref{w22}) by means of a Lagrange multiplier $k(x)$. Next, 
since the current ${\bf j}_\mu^b$ describes a magnetically charged 
excitation of the vacuum, it is directed along some of the  
root vectors, ${\bf q}_b$, where ``$b$'' is a certain fixed index from the set 
$1,\ldots,\frac{N_c(N_c-1)}{2}$. Therefore, one 
can write ${\bf j}_\mu^b={\bf q}_b j_\mu$. Introducing auxiliary fields $C_\mu^a$'s as 
$${\rm e}^{-\eta^2\int_x\sum\limits_a(\partial_\mu\theta_a+\mu 
{\bf q}_a{\bf v}_\mu)^2}=\int \prod\limits_a {\cal D}C_\mu^a{\,} {\rm e}^{\int_x\left[-\frac{1}{4\eta^2}(C_\mu^a)^2+
iC_\mu^a (\partial_\mu\theta_a+\mu 
{\bf q}_a{\bf v}_\mu)\right]},$$
one obtains, similarly to the 4-d case considered in Refs.~\cite{ae,s3}, the following result:
$$\int\left(\prod\limits_a {\cal D}\tilde\theta_a{\,}{\cal D}\bar\theta_a\right)
{\cal D}k{\,}\delta\left(\sum\limits_{a}\tilde\theta_a\right){\rm e}^{-\int_x\left[\eta^2\sum\limits_a
(\partial_\mu\theta_a+\mu 
{\bf q}_a{\bf v}_\mu)^2-ik\sum\limits_a\bar\theta_a\right]}=$$
$$=\int\left(\prod\limits_a {\cal D} J_\mu^a{\,} {\cal D}\varphi_\mu^a\right)\delta\left(\sum\limits_a
J_\mu^a\right){\rm e}^{\int_x\left[-\frac{1}{8\eta^2}(\Phi_{\mu\nu}^a)^2+i\mu\varepsilon_{\mu\nu\lambda}
{\bf q}_a{\bf v}_\mu\partial_\nu\varphi_\lambda^a+2\pi i\varphi_\mu^aJ_\mu^a\right]}.$$
Here, it has been taken into account 
that $\sum\limits_{a}{\bf q}_a=0$, owing to which the 
$k$-integration yields just an inessential global normalization constant. Furthermore, the constraint 
$\sum\limits_{a}^{}\tilde\theta_a=0$ went over into 
$\sum\limits_a J_\mu^a=0$, which means that the theory actually contains $\frac{N_c(N_c-1)}{2}-1$ types of 
mutually independent vortices (cf. Refs.~\cite{ae,s3} for a similar constraint for the dual strings).

To further perform the integration over ${\bf v}_\mu$, it is convenient to introduce 
the fields $u_\mu^a={\bf q}_a{\bf v}_\mu$, and use the formula~\cite{s3, s4} $\sum\limits_{a}q_a^\alpha q_a^\beta=\frac{N_c}{2}\delta^{\alpha\beta}$. Recalling that ${\bf j}_\mu^b={\bf q}_b j_\mu$, we can then represent 
the ${\bf v}_\mu$-dependent part of the action as
$$\int_x\left[\frac{1}{4\nu}{\bf F}_{\mu\nu}^2-i{\bf v}_\mu\left(
\mu\varepsilon_{\mu\nu\lambda}{\bf q}_a\partial_\nu\varphi_\lambda^a+\frac{1}{\nu}{\bf j}_\mu^b\right)\right]=
\int_x\left[\frac{1}{2N_c\nu}\left(\partial_\mu u_\nu^a-\partial_\nu u_\mu^a\right)^2-iu_\mu^aK_\mu^a\right],$$
where $K_\mu^a=
\mu\varepsilon_{\mu\nu\lambda}\partial_\nu\varphi_\lambda^a+\frac{1}{\nu}\delta^{ab}j_\mu$. 
Then the Gaussian integration over $u_\mu^a$'s readily yields the action 
$\frac{N_c\nu}{4}\int_{x,y} K_\mu^{a,x}D_0^{xy}K_\mu^{a,y}$, which can be further simplified by 
representing $K_\mu^a$ as 
$K_\mu^a=\varepsilon_{\mu\nu\lambda}\partial_\nu\left(\mu\varphi_\lambda^a+\frac{1}{\nu}\delta^{ab}
\varepsilon_{\lambda\alpha\beta}\int_y\partial_\alpha^x D_0^{xy}j_\beta^y\right)$. In this way, 
we obtain the following ($N_c>2$)-counterpart of Eq.~(\ref{w5}):
$$\left<W_b(C)\right>=$$
$$=\int\left(\prod\limits_a {\cal D} J_\mu^a{\,} {\cal D}\varphi_\mu^a\right)\delta\left(\sum\limits_a
J_\mu^a\right){\rm e}^{\int_x\left[-\frac{1}{8\eta^2}(\Phi_{\mu\nu}^a)^2+2\pi i\varphi_\mu^a J_\mu^a-
\frac{N_c\nu}{4}\left(\mu\varphi_\mu^a+\frac1\nu\delta^{ab}\varepsilon_{\mu\nu\lambda}\int_y\partial_\nu^x
D_0^{xy}j_\lambda^y\right)^2\right]}.$$
This expression can finally be brought to the form similar to that of Eq.~(\ref{a2}). Indeed, 
proceeding in the same way as from Eq.~(\ref{w5}) to Eq.~(\ref{a2}),
we obtain the following final result:
$$
\left<W_b(C)\right>=$$
\be
\label{v7}
={\rm e}^{-\frac{N_c}{4\nu}\int_{x,y}j_\mu^xj_\mu^yD_{\sf m}^{xy}}
\int\prod\limits_a {\cal D} J_\mu^a{\,}\delta\left(\sum\limits_a
J_\mu^a\right)
{\rm e}^{-(2\pi\eta)^2\int_{x,y}J_\mu^{a,x}J_\mu^{a,y}D_{\sf m}^{xy}+
\frac{2\pi i}{\mu\nu}\left[\hat L(j,J^b)-\varepsilon_{\mu\nu\lambda}
\int_{x,y}J_\mu^{b,x} j_\nu^y\partial_\lambda^xD_{\sf m}^{xy}\right]},
\ee
where ${\sf m}=\mu\eta\sqrt{N_c\nu}$ generalizes Eq.~(\ref{mml}) for the mass of the dual vector boson. 
Thus, Eq.~(\ref{v7}) represents the sought generalization of Eq.~(\ref{a2}) to the 
case of $N_c>2$. We notice that, while the strength of the $(j\times j)$-interaction becomes $(N_c/2)$ times larger
compared to that of Eq.~(\ref{a2}),
the coefficient at the linking number remains the same. Therefore, much as in the 
SU(2)-inspired case, in the general 
SU($N_c$)-inspired model considered in this Section, the Aharonov--Bohm-type interaction between the 
integer-charged excitations of the dual Higgs vacuum and the 
dual Abrikosov vortices yields only a trivial factor of ${\rm e}^{2\pi i\times({\rm integer})}$.

\section{Dual Wilson loop and its interaction with Abrikosov vortices in the presence of a Chern--Simons term}

We extend now the analysis performed in Section~II to the case where the CS term is included. 
This term is known to produce self-linkage of the contour of a Wilson loop~\cite{5}, and we expect 
that it would lead to a similar effect for the dual Abrikosov vortices.
To start with, we 
again consider the theory where the dual Higgs field is absent, that is equivalent to setting 
$\eta=0$. The Wilson loop in such a theory is given by the following extension of Eq.~(\ref{1}): 
$$\left<W(C)\right>=\int {\cal D}v_\mu{\,}
{\rm e}^{-\int_x\left[\frac{1}{4\nu}F_{\mu\nu}^2[v]+
i\Theta\varepsilon_{\mu\nu\lambda}v_\mu\partial_\nu v_\lambda
-\frac{i}{\nu}v_\mu j_\mu\right]},$$
where the dimensionality of the new parameter $\Theta$ is (mass)$^2$. 
Imposing the 
gauge-fixing condition $\partial_\mu v_\mu=0$, we obtain the saddle-point equation 
$$-\partial^2v_\mu+im\varepsilon_{\mu\nu\lambda}\partial_\nu v_\lambda=ij_\mu,~~~~  {\rm where}~~~~ 
m=2\Theta\nu.$$
Seeking a solution in the form $v_\mu=U_\mu+iV_\mu$, we get a system of equations
\begin{equation}
\label{nm55}
\partial^2U_\mu+m\varepsilon_{\mu\nu\lambda}\partial_\nu V_\lambda=0,~~~~~~ 
-\partial^2V_\mu+m\varepsilon_{\mu\nu\lambda}\partial_\nu U_\lambda=j_\mu.
\end{equation}
The first of these equations can be solved with respect to $U_\mu$ as
\begin{equation}
\label{amu}
U_\mu^x=m\varepsilon_{\mu\nu\lambda}\int_y D_0^{xy}\partial_\nu^y V_\lambda^y.
\end{equation}
Differentiating the second equation~(\ref{nm55}), and applying the maximum principle, one gets
$\partial_\mu V_\mu=0$. Using this relation,
one further obtains from Eq.~(\ref{amu}): $\varepsilon_{\mu\nu\lambda}\partial_\nu U_\lambda=mV_\mu$. 
The substitution of this formula into the 
second equation~(\ref{nm55}) yields for that equation a remarkably simple form $(-\partial^2+m^2)V_\mu=j_\mu$.
Therefore, one has $V_\mu^x=\int_y D_m^{xy}j_\mu^y$, while $U_\mu^x$, given by Eq.~(\ref{amu}), can be calculated by virtue of Eq.~(\ref{a23}), and reads
$U_\mu^x=\frac1m\varepsilon_{\mu\nu\lambda}\int_y(D_0^{xy}-D_m^{xy})\partial_\nu^y j_\lambda^y$.
Altogether, the resulting Wilson loop has the form 
\begin{equation}
\label{dl8}
\left<W(C)\right>\bigr|_{\eta=0}=\exp\left\{\frac{1}{2\nu}\int_{x,y}\left[-j_\mu^xD_m^{xy}j_\mu^y+
\frac{i}{m}\varepsilon_{\mu\nu\lambda}j_\mu^x j_\lambda^y\partial_\nu^x(D_0^{xy}-D_m^{xy})\right]\right\}.
\end{equation}
Recalling the definition of the parameter $\nu$ from Eq.~(\ref{numu}), we observe that the obtained Eq.~(\ref{dl8})
extends Eq.~(\ref{wt}) to the case of $\Theta\ne 0$. Clearly, the $\Theta$-term 
leads to a self-linkage of the contour $C$, as well as to a short-range self-interaction
of this contour by means of the Yukawa propagator $D_m^{xy}$. We also notice that,  
when $\Theta\to 0$ in Eq.~(\ref{dl8}), one recovers Eq.~(\ref{wt}). Indeed, in this limit, 
one has $\frac{1}{m}(D_0^{xy}-D_m^{xy})\to\frac{1}{4\pi}$, so that 
$$\frac1m\int_{x,y}j_\mu^x j_\lambda^y\partial_\nu^x(D_0^{xy}-D_m^{xy})=
\frac1m\int_{x,y}j_\mu^x (D_0^{xy}-D_m^{xy})\partial_\nu^y j_\lambda^y\to\frac{1}{4\pi}\int_{x,y}
j_\mu^x\partial_\nu^y j_\lambda^y=0,$$
since $\int_x j_\mu^x=0$.

We proceed now to the duality transformation of the Wilson loop in the full theory, where the dual Higgs field 
is present and its condensation does take place, i.e. $\eta\ne 0$. 
The corresponding extension of Eq.~(\ref{1}) reads
\begin{equation}
\label{in}
\left<W(C)\right>=\int {\cal D}v_\mu{\,}{\cal D}\tilde\theta{\,}{\cal D}\bar\theta{\,}
{\rm e}^{-\int_x\left[\frac{1}{4\nu}F_{\mu\nu}^2[v]+\eta^2(\partial_\mu\theta+\mu v_\mu)^2+
i\Theta\varepsilon_{\mu\nu\lambda}v_\mu\partial_\nu v_\lambda
-\frac{i}{\nu}v_\mu j_\mu\right]}.
\end{equation}
The transformation leading from Eq.~(\ref{1}) to Eq.~(\ref{qq}) remains the same, so that the 
counterpart of Eq.~(\ref{qq}) in the presence of the CS term has the form
$$\left<W(C)\right>=\int {\cal D}J_\mu{\,}{\cal D}\varphi_\mu{\,}{\cal D}G_\mu{\,}
{\cal D}v_\mu{\,}
{\rm e}^{\int_x\left\{-\frac{\nu}{2}G_\mu^2-\frac{1}{8\eta^2}\Phi_{\mu\nu}^2+
i v_\mu\left[\varepsilon_{\mu\nu\lambda}\partial_\nu(G_\lambda+\mu\varphi_\lambda-\Theta v_\lambda)+
\frac1\nu j_\mu\right]
+2\pi i\varphi_\mu J_\mu\right\}}.$$
Unlike the case where the CS term was absent, the field 
$v_\mu$ now ceases to be a Lagrange multiplier. Nevertheless, since the $v_\mu$-integration 
is Gaussian, it can be performed exactly, and we proceed to this integration.

The corresponding saddle-point equation for $v_\mu$ reads
$\varepsilon_{\mu\nu\lambda}\partial_\nu v_\lambda=\frac{1}{2\Theta}k_\mu$, where we have denoted
$k_\mu=\varepsilon_{\mu\nu\lambda}\partial_\nu(G_\lambda+\mu\varphi_\lambda)+\frac1\nu j_\mu$.
Owing to the conservation of $k_\mu$, a solution to this saddle-point equation reads $v_\mu^x=\frac{1}{2\Theta}
\varepsilon_{\mu\nu\lambda}\partial_\nu^x\int_y D_0^{xy}k_\lambda^y$. Plugging this solution back into the 
exponent ${\rm e}^{i\int_x v_\mu(k_\mu-\Theta\varepsilon_{\mu\nu\lambda}\partial_\nu v_\lambda)}$, and using the 
above explicit expression for $k_\mu$, we obtain, upon some algebra, the following formula:
$$\left<W(C)\right>={\rm e}^{\frac{i}{2\nu m}\varepsilon_{\mu\nu\lambda}\int_{x,y}j_\mu^x j_\lambda^y
\partial_\nu^xD_0^{xy}}\times$$
\begin{equation}
\label{yy}
\times\int {\cal D}J_\mu{\,}{\cal D}\varphi_\mu{\,}{\cal D}G_\mu{\,}
{\rm e}^{\int_x\left\{-\frac{\nu}{2}G_\mu^2-\frac{1}{8\eta^2}\Phi_{\mu\nu}^2+
\frac{i}{4\Theta}\varepsilon_{\mu\nu\lambda}\left[G_\mu\partial_\nu(G_\lambda+2\mu\varphi_\lambda)+
\mu^2\varphi_\mu\partial_\nu\varphi_\lambda\right]+\frac{i}{2\Theta\nu}(G_\mu+\mu\varphi_\mu)j_\mu
+2\pi i\varphi_\mu J_\mu\right\}}.
\end{equation}
Here, the argument of the first exponent coincides with the term containing 
the Gauss' self-linking number of the contour $C$, which was present already in Eq.~(\ref{dl8}). 
In addition, the functional integral in Eq.~(\ref{yy}) 
describes interactions of the dual-Higgs excitation with the dual Abrikosov vortices, 
as well as their self-interactions in the presence of the CS term.

In order to visualize all these interactions, let us perform the $G_\mu$-integration first. 
Representing the saddle-point expression for $G_\mu$ in the 
form $G_\mu=L_\mu+iN_\mu$, we obtain a system of two saddle-point equations:
$$\varepsilon_{\mu\nu\lambda}\partial_\nu L_\lambda-mN_\mu+n_\mu=0,~~~~~~
\varepsilon_{\mu\nu\lambda}\partial_\nu N_\lambda+mL_\mu=0,$$
where we have denoted $n_\mu=\mu\varepsilon_{\mu\nu\lambda}\partial_\nu \varphi_\lambda+\frac1\nu j_\mu$. Owing to the conservation of $n_\mu$, we find a solution
to these equations in the form
$$L_\mu^x=-\varepsilon_{\mu\nu\lambda}\int_y D_m^{xy}\partial_\nu^y n_\lambda^y,~~~~~~
N_\mu^x=m\int_y D_m^{xy}n_\mu^y.$$
Plugging the corresponding saddle-point expression for $G_\mu$ back into Eq.~(\ref{yy}), we obtain, after some algebra, the following general result:
$$\int {\cal D}G_\mu{\,}
{\rm e}^{\int_x\left(-\frac{\nu}{2}G_\mu^2+\frac{i}{4\Theta}\varepsilon_{\mu\nu\lambda}G_\mu\partial_\nu
G_\lambda+\frac{i}{2\Theta}G_\mu k_\mu\right)}=$$
$$={\rm e}^{-\frac{1}{2\nu}\int_{x,y}j_\mu^x j_\mu^y D_m^{xy}
-\mu\varepsilon_{\mu\nu\lambda}\int_{x,y}
D_m^{xy}j_\mu^x\partial_\nu^y\varphi_\lambda^y
+\frac{\nu\mu^2}{2}\left[\int_{x,y}D_m^{xy}\cdot\left(
m^2\varphi_\mu^x\varphi_\mu^y+\partial_\mu^x\varphi_\mu^x\cdot\partial_\nu^y\varphi_\nu^y\right)-
\int_x \varphi_\mu^2\right]}\times$$
\begin{equation}
\label{5s}
\times {\rm e}^{-\frac{i}{4\Theta}\left\{\mu^2\varepsilon_{\mu\nu\lambda}\int_x\varphi_\mu
\partial_\nu\varphi_\lambda+\varepsilon_{\mu\nu\lambda}\int_{x,y}D_m^{xy}\cdot\left[\frac{1}{\nu^2}
j_\mu^x\partial_\nu^y j_\lambda^y-(\mu m)^2\varphi_\mu^x\partial_\nu^y\varphi_\lambda^y\right]+
\frac{2\mu}{\nu}\left(\int_x \varphi_\mu j_\mu-m^2\int_{x,y}D_m^{xy}\varphi_\mu^x j_\mu^y\right)\right\}}.
\end{equation}
We notice that, 
in the limit of $\nu\to 0$, the initial Eq.~(\ref{in}) yields Eq.~(\ref{wt}):
\begin{equation}
\label{hh}
\left<W(C)\right>\rightarrow\int {\cal D}v_\mu{\,}
{\rm e}^{-\int_x\left(\frac{1}{4\nu}F_{\mu\nu}^2
-\frac{i}{\nu}v_\mu j_\mu\right)}={\rm e}^{-\frac{1}{2\nu}\int_{x,y}j_\mu^x j_\mu^y D_0^{xy}}.
\end{equation}
Therefore, the remaining 
$\varphi_\mu$-integration in Eq.~(\ref{yy}) should also yield Eq.~(\ref{wt}) in this limit.
The limit of $\nu\to 0$ can thus serve as a check for Eq.~(\ref{5s}). 
The right-hand side of Eq.~(\ref{5s}) simplifies in this limit to the form
$${\rm e}^{-\frac{1}{2\nu}\int_{x,y}j_\mu^x j_\mu^y D_0^{xy}
-\mu\varepsilon_{\mu\nu\lambda}\int_{x,y}
D_0^{xy}j_\mu^x\partial_\nu^y\varphi_\lambda^y-\frac{i}{4\Theta}\left(\mu^2\varepsilon_{\mu\nu\lambda}
\int_x\varphi_\mu\partial_\nu\varphi_\lambda+\frac{1}{\nu^2}\varepsilon_{\mu\nu\lambda}\int_{x,y}D_0^{xy}
j_\mu^x\partial_\nu^y j_\lambda^y+\frac{2\mu}{\nu}\int_x \varphi_\mu j_\mu\right)},$$
and the Wilson loop~(\ref{yy}) becomes
$$\left<W(C)\right>\rightarrow
{\rm e}^{-\frac{1}{2\nu}\int_{x,y} j_\mu^x j_\mu^y D_0^{xy}
+\frac{i}{4\Theta\nu^2}\varepsilon_{\mu\nu\lambda}\int_{x,y}\left(j_\mu^x j_\lambda^y\partial_\nu^x
D_0^{xy}-D_0^{xy}j_\mu^x\partial_\nu^y j_\lambda^y\right)}\times$$
$$\times
\int {\cal D}J_\mu{\,}{\cal D}\varphi_\mu{\,}{\rm e}^{\int_x\left(-\frac{1}{8\eta^2}\Phi_{\mu\nu}^2+
2\pi i\varphi_\mu J_\mu\right)-\mu\varepsilon_{\mu\nu\lambda}\int_{x,y}\varphi_\mu^x j_\lambda^y
\partial_\nu^x D_0^{xy}}.$$
The Gaussian $\varphi_\mu$-integration in this formula yields, upon some algebra,
$$\left<W(C)\right>\rightarrow
{\rm e}^{-\frac{1}{2\nu}\int_{x,y} j_\mu^x j_\mu^y D_0^{xy}}
\int {\cal D}J_\mu{\,}{\rm e}^{-(2\pi\eta)^2\int_{x,y}J_\mu^x J_\mu^y D_0^{xy}}.
$$
Recalling the normalization of the integration measure ${\cal D}J_\mu$, discussed in Appendix~A,
we indeed recover the expected result~(\ref{hh}). Thus, our check of Eq.~(\ref{5s}) was successful.

We consider now large values of the $\Theta$-parameter, namely such that 
\begin{equation}
\label{ineq2}
\Theta\gg\kappa\mu\eta.
\end{equation} 
According to Eq.~(\ref{numu}),
such large $\Theta$'s imply $m\gg\kappa\eta$, that makes the action in the exponentials on the right-hand side of Eq.~(\ref{5s}) local, and brings 
the Wilson loop to the form 
$$\left<W(C)\right>\rightarrow{\rm e}^{-\frac{1}{2\nu m^2}\int_x j_\mu^2+\frac{i}{4\Theta\nu^2}
\varepsilon_{\mu\nu\lambda}\left(\int_{x,y}j_\mu^x j_\lambda^y\partial_\nu^x D_0^{xy}-
\frac{1}{m^2}\int_x j_\mu
\partial_\nu j_\lambda\right)}\times$$
\begin{equation}
\label{n3}
\times\int{\cal D}J_\mu{\,}{\cal D}\varphi_\mu{\,}{\rm e}^{\int_x\left[-\frac{1}{8\eta^2}\Phi_{\mu\nu}^2+
\frac{\mu^2}{8\Theta^2\nu}(\partial_\mu\varphi_\mu)^2+\frac{i\mu^2}{4\Theta}\varepsilon_{\mu\nu\lambda}
\varphi_\mu\partial_\nu\varphi_\lambda+i\varphi_\mu\left(2\pi J_\mu+\frac{\mu}{m}j_\mu+
\frac{i\mu}{m^2}\varepsilon_{\mu\nu\lambda}\partial_\nu j_\lambda\right)\right]}.
\end{equation}
Furthermore, in the 
same limiting case~(\ref{ineq2}), the $\varphi_\mu$-integration in this formula can also be performed analytically. Referring the reader for the details to Appendix~B, we present here 
the final result of this integration:
$$\left<W(C)\right>\rightarrow{\rm e}^{-\frac{1}{2\nu m^2}\int_x j_\mu^2-\frac{i\Theta}{m^4}
\varepsilon_{\mu\nu\lambda}\int_x j_\mu\partial_\nu j_\lambda}\times$$
\begin{equation}
\label{rr}
\times\int {\cal D}J_\mu{\,}{\rm e}^{-\eta^2\int_{x,y}R_\mu^x R_\mu^y D_{\cal M}^{xy}+\frac{i\Theta}{\mu^2}
\varepsilon_{\mu\nu\lambda}\int_{x,y}\left[R_\mu^x R_\lambda^y\partial_\nu^x D_{\cal M}^{xy}-4\pi J_\mu^x
\left(\pi J_\lambda^y+\frac{\mu}{m} j_\lambda^y\right)\partial_\nu^x D_0^{xy}\right]}.
\end{equation}
In this formula, ${\cal M}\equiv\frac{\mu^2\eta^2}{\Theta}$, and $R_\mu\equiv 2\pi J_\mu+\frac{\mu}{m}j_\mu$.
Remarkably, in the limit~(\ref{ineq2}), the initial CS term for the velocity, $i\Theta\varepsilon_{\mu\nu\lambda}v_\mu\partial_\nu v_\lambda$ from Eq.~(\ref{in}), 
leads to the appearance of its counterpart 
$\frac{i\Theta}{m^4}
\varepsilon_{\mu\nu\lambda} j_\mu\partial_\nu j_\lambda$
for the current $j_\mu$, while 
the self-linkage of the contour $C$, described by the first exponential in Eq.~(\ref{yy}), disappears. Rather, we observe the appearance of a self-linkage of the 
dual Abrikosov vortices, as well as of their linkage with the contour $C$, as described by the term
$\frac{4\pi i\Theta}{\mu^2}\varepsilon_{\mu\nu\lambda} J_\mu^x
\left(\pi J_\lambda^y+\frac{\mu}{m} j_\lambda^y\right)\partial_\nu^x D_0^{xy}$ in the Lagrangian. In particular,
the part $\frac{4\pi i\Theta}{\mu^2}\varepsilon_{\mu\nu\lambda} J_\mu^x\cdot\frac{\mu}{m} j_\lambda^y
\partial_\nu^x D_0^{xy}$ of this expression yields in the action the same term $-\frac{2\pi i}{\mu\nu}\hat L(j,J)$
as in the absence of the CS term (cf. the end of Section~II). 
Thus, in the presence of the CS term, the Aharonov--Bohm-type interaction of the dual-Higgs excitation 
with the dual Abrikosov vortex gets trivial in the limit~(\ref{ineq2}).
Rather, the term $\frac{4\pi^2i\Theta}{\mu^2}\varepsilon_{\mu\nu\lambda}J_\mu^xJ_\lambda^y\partial_\nu^xD_0^{xy}$
means that the CS term makes dual Abrikosov vortices knotted as long as the condition  $$\frac{\mu^2}{\Theta}\ne\frac{2\pi}{\rm integer}$$
is met, where the parameter $\mu$ is defined in Eq.~(\ref{numu}).

\section{Summary}

The spatial confinement in the dimensionally-reduced high-temperature gluodynamics can be modelled by means 
of the dual Landau--Ginzburg-type theory. 
In this paper, we have explored interactions between an excitation of the dual-Higgs vacuum
and the dual Abrikosov vortices, which are present in such a theory. 
For this purpose, starting with the simplest SU(2)-inspired case, 
we have performed a duality transformation of the corresponding Wilson loop~(\ref{we}).
The resulting Eq.~(\ref{a2}) contains 
a long-range Aharonov--Bohm-type interaction of the dual-Higgs excitation with the dual Abrikosov vortices,
which is represented by the Gauss' linking number. However, we have found the coefficient at this linking number to
be $2\pi i\times({\rm integer})$, which makes the said Aharonov--Bohm-type interaction trivial.
In Section~III, we have obtained the same trivialization for the case of 
the SU($N_c$)-inspired dual Landau--Ginzburg-type theory,
and in Section~IV --- at the sufficiently large values of the $\Theta$-parameter in the theory extended 
by the CS term. Thus, in all these cases, massless interactions drop out altogether from the dual formulation 
of the Wilson loop, so that the interactions between the dual-Higgs excitation and the dual Abrikosov vortices 
are mediated entirely by the dual vector bosons. Finally, we have explicitly demonstrated a qualitatively novel 
phenomenon of the appearance of knotted dual Abrikosov vortices due to the CS term.

\begin{acknowledgments}

\noindent
This work was supported by the Portuguese Foundation for Science and Technology
(FCT, program Ci\^encia-2008) and by 
the Center for Physics of Fundamental Interactions (CFIF) at Instituto Superior
T\'ecnico (IST), Lisbon. The author is grateful to the whole staff of the Department of Physics of IST
for their cordial hospitality. 
\end{acknowledgments}

\appendix

\section{Some details of the derivation of Eq.~(\ref{a2})}

With the use of Eq.~(\ref{k33}), and owing to the conservation of 
$j_\mu$, one has
$$-\eta^2\int_{x,y}K_\mu^xK_\mu^y D_{\sf m}^{xy}=-(2\pi\eta)^2\int_{x,y}J_\mu^xJ_\mu^yD_{\sf m}^{xy}-
4\pi i\mu\eta^2\varepsilon_{\mu\nu\lambda}\int_{x,y}D_{\sf m}^{xy}J_\mu^x\partial_\nu^y\int_z D_0^{yz}j_\lambda^z+$$
\begin{equation}
\label{aa}
+(\mu\eta)^2\int_{x,y}D_{\sf m}^{xy}\left(\partial_\nu^x\int_z D_0^{xz}j_\lambda^z\right)\left(\partial_\nu^y
\int_u D_0^{yu}j_\lambda^u\right).
\end{equation}
We furthermore assume the standard normalization $\left<1\right>=1$ of the functional average, which implies 
a division by the functional integral 
$\int {\cal D}J_\mu{\,}{\rm e}^{-(2\pi\eta)^2\int_{x,y}J_\mu^xJ_\mu^yD_{\sf m}^{xy}}$ corresponding to the 
first term on the right-hand side of Eq.~(\ref{aa}). Thus, we always imply that the 
measure ${\cal D}J_\mu$ is normalized by a division by this integral.

The last term in Eq.~(\ref{aa}) can be represented, through the integration by parts, as $(\mu\eta)^2\int_{x,y}
D_{\sf m}^{xy}j_\mu^x\int_u D_0^{yu}j_\mu^u$.
The $y$-integration in this expression is straightforward:
\begin{equation}
\label{a23}
\int_y D_{\sf m}^{xy}D_0^{yu}=\int_y\int_p\frac{{\rm e}^{i{\bf p}({\bf x}-{\bf y})}}{{\bf p}^2+{\sf m}^2}
\int_q\frac{{\rm e}^{i{\bf q}({\bf y}-{\bf u})}}{{\bf q}^2}=
\int_p\frac{{\rm e}^{i{\bf p}({\bf x}-{\bf u})}}{{\bf p}^2
({\bf p}^2+{\sf m}^2)}=\frac{1}{{\sf m}^2}\left(D_0^{xu}-D_{\sf m}^{xu}\right),
\end{equation}
where the equality $\frac{1}{{\bf p}^2
({\bf p}^2+{\sf m}^2)}=\frac{1}{{\sf m}^2}\left(\frac{1}{{\bf p}^2}-\frac{1}{{\bf p}^2+{\sf m}^2}\right)$ has been used
at the last step. Using further the explicit form of ${\sf m}$, Eq.~(\ref{mml}), we can represent the last term in 
Eq.~(\ref{aa}) as 
$\frac{1}{2\nu}\int_{x,y}j_\mu^x j_\mu^y
\left(D_0^{xy}-D_{\sf m}^{xy}\right)$.

In the second term on the right-hand side of Eq.~(\ref{aa}), one can use the equality 
$\partial_\nu^y\int_z D_0^{yz}j_\lambda^z=\int_z D_0^{yz}\partial_\nu^z j_\lambda^z$, which yields the same $y$-integration as in Eq.~(\ref{a23}): $\int_y D_{\sf m}^{xy}D_0^{yz}=
\frac{1}{{\sf m}^2}\left(D_0^{xz}-D_{\sf m}^{xz}\right)$. 
Upon the subsequent integration by 
parts, we obtain for this term the following expression: $\frac{2\pi i}{\mu\nu}\varepsilon_{\mu\nu\lambda}
\int_{x,y}J_\mu^x j_\nu^y\partial_\lambda^x\left(D_0^{xy}-D_{\sf m}^{xy}\right)$. Noticing also the definition 
of the Gauss' linking number, 
$\hat L(j,J)=\varepsilon_{\mu\nu\lambda}\int_{x,y}J_\mu^xj_\nu^y\partial_\lambda^xD_0^{xy}$, 
we arrive at Eq.~(\ref{a2}).

\section{Some details of the derivation of Eq.~(\ref{rr})}

For $\Theta$'s obeying condition~(\ref{ineq2}), one readily obtains the inequality 
\begin{equation}
\label{in33}
\frac{\mu^2}{\Theta^2\nu}\ll\frac{1}{\eta^2}.
\end{equation}
Owing to this inequality, the term 
$\frac{\mu^2}{8\Theta^2\nu}(\partial_\mu\varphi_\mu)^2$ in Eq.~(\ref{n3}) can be neglected 
in comparison with the absolute value of the term
$-\frac{1}{8\eta^2}\Phi_{\mu\nu}^2$. The resulting Gaussian $\varphi_\mu$-integration can be performed by  seeking the saddle-point function in the form $\varphi_\mu=\varphi_\mu^{(1)}+i\varphi_\mu^{(2)}$, and solving 
the so-emerging system of equations for $\varphi_\mu^{(1)}$ and $\varphi_\mu^{(2)}$.
The result can be written as 
\begin{equation}
\label{d5}
\int {\cal D}\varphi_\mu{\,}{\rm e}^{\int_x\left[\cdots\right]}={\rm e}^{\frac12\int_x
\left[-R_\mu\varphi_\mu^{(2)}-S_\mu\varphi_\mu^{(1)}+i\left(R_\mu\varphi_\mu^{(1)}-
S_\mu\varphi_\mu^{(2)}\right)\right]},
\end{equation}
where $R_\mu\equiv 2\pi J_\mu+\frac{\mu}{m}j_\mu$ and $S_\mu\equiv\frac{\mu}{m^2}
\varepsilon_{\mu\nu\lambda}\partial_\nu j_\lambda$ are respectively 
the real and the imaginary parts of the current 
which couples to $\varphi_\mu$ in Eq.~(\ref{n3}). The obtained real 
and imaginary parts of the saddle-point function $\varphi_\mu$ entering Eq.~(\ref{d5}) read
$$\varphi_\mu^{(1)}=$$
\begin{equation}
\label{ph1}
=2\mu\eta^2\varepsilon_{\mu\nu\lambda}\left\{\frac{\mu\eta^2}{\Theta {\cal M}^2}\int_y
\left[2\pi J_\lambda^y+\frac{\mu}{m}\left(1-\frac{\cal M}{m}\right)j_\lambda^y\right]\partial_\nu^x\left(
D_{\cal M}^{xy}-D_0^{xy}\right)-\frac{1}{m^2}\int_y j_\lambda^y\partial_\nu^x D_0^{xy}\right\}
\end{equation}
and
\begin{equation}
\label{ph2}
\varphi_\mu^{(2)}=2\eta^2\int_y D_{\cal M}^{xy}\left[2\pi J_\mu^y+\frac{\mu}{m}\left(1-
\frac{\cal M}{m}\right)j_\mu^y
\right],
\end{equation}
with the new mass parameter ${\cal M}\equiv\frac{\mu^2\eta^2}{\Theta}$. Furthermore, in the limit~(\ref{in33}) at issue,
the ${\cal O}({\cal M}/m)$-terms in Eqs.~(\ref{ph1}) and (\ref{ph2})
should be neglected compared to 1.
That yields the following saddle-point expressions for $\varphi_\mu^{(1)}$ and $\varphi_\mu^{(2)}$:
$$\varphi_\mu^{(1)}=\frac{2\Theta}{\mu^2}\varepsilon_{\mu\nu\lambda}\int_y R_\lambda^y{\,}
\partial_\nu^x\left(D_{\cal M}^{xy}-D_0^{xy}\right),~~~ 
\varphi_\mu^{(2)}=2\eta^2\int_y R_\mu^y{\,} D_{\cal M}^{xy}.$$
Substituting them into Eq.~(\ref{d5}), we obtain for the Wilson loop in the limit~(\ref{ineq2}) 
expression~(\ref{rr}).

\end{document}